# Electric cars – assessment of 'green' nature vis-à-vis conventional fuel driven cars

Satish Vitta[*]

Department of Metallurgical Engineering and Materials Science

Indian Institute of Technology Bombay

Mumbai 400076; India.

## Abstract

A comprehensive analysis of energy requirements and emissions associated with electric vehicles, ranging from mining and making the rare-earth magnets required in electric motor to assembling the Li-ion battery, including charging and regular running of the electric vehicles has been performed. A simple, analytical procedure is used to determine the embodied energy and emissions. The objective is to assess the potential of electric cars to reduce green house gases emission to limit global warming to < 1.5 °C by the Year 2050 as per IPCC recommendations and also to compare them with conventional fuel driven cars. The combined embodied energy for Nd- and Dy-metals production which are required in electric motors and battery assembly for 150 million cars, projected to be on the road in the year 2050 is ~ 1500 TWh and the $CO_2$ emissions is found to be > 600 MT. The emissions includes carbon intensity of electrical energy required to run these electric vehicles. The projected emissions due to fossil fuels, gasoline production as well as burning it in combustion engines however is only 412 MT, far less than that due to electric vehicles. The main contributor to emissions from electric vehicles is the battery assembling process which releases ~ 379 MT of $CO_2$-e gases. The emissions from both electric vehicles as well as combustion engine vehicles scale linearly with the number of vehicles, indicating that a breakeven is not possible with the currently available manufacturing technologies. These results clearly show that significant technological developments have to take place in electric vehicles so that they become environmentally better placed compared to combustion engine based cars.

* Email: satish.vitta@iitb.ac.in

**Introduction:**

The most versatile form of energy is 'Electricity' which is used for a host of applications from house hold utilities to large scale transportation. Another versatile form of energy which is used for a lot of activities is mechanical energy. Both these forms of energy are not naturally occurring and can only be obtained by converting naturally occurring resources of energy: fossil fuels, solar energy, hydro & hydrothermal energy, wind energy and nuclear energy. All the known energy conversion technologies however have limited conversion efficiencies and more importantly they result in generation of global warming greenhouse gases, GHG such as $CO_2$, $NO_x$, $SO_x$, $CH_4$, CFC and so on. These result in warming and raise the 'Global Mean Surface Temperature', which has been rising steadily at the rate of ~ 0.2° ± 0.1 °C per decade, reading 1.0° C above the pre-industrial value in 2019.[1] The global $CO_2$ emissions alone are ~ 36.3 Gt in the Year 2019[2] with contributions from other GHG not included in this estimate. The main contributor to these emissions is electricity and heat generation with transportation contributing up to 25 % of these emissions depending on geographical location. Transportation and electricity production have several common features/modes across nations and hence feasibility of technological interventions to yield verifiable results is very high. As a result these two sectors are being prioritized for decarbonisation and reducing GHG. Electricity production is being shifted to renewable resources such as solar energy conversion and harnessing wind energy. Transportation on the other hand is being shifted from conventional oil based internal combustion engine driven vehicles to electric motor operated vehicles and is anticipated to decarbonize transportation. As a result most of the countries are adopting these technologies through various legislations and hence all the current automobile manufacturers are driven to develop a variety of electric vehicles and release them every year. The flip side of this development is that implementation of these technologies puts constraints on certain materials which are becoming critical. Also, production of these materials which are critical to implement these technologies have an embodied energy and environmental costs which are not insignificant and certainly need to be factored into. Added to this embodied energy cost of the various materials is the geo-political consideration of materials supply chain risk, generally assessed using the Herfindahl-Hirschman Index. There are several studies in the literature which mainly address issues concerning Li-ion batteries and their recycling necessity including their embodied energy.[3-5] The materials requirement of electric motors together with their extraction and processing however have not been discussed.[6] A single study which consolidates all aspects of electric vehicles – motors to batteries to operation of these electric vehicles has not been done. Hence the main objective of this study has been to estimate the different materials requirement from simple first principles, materials intensity and to critically analyse this intensity in terms of availability, including the emissions that result during their production. The highlighting feature of this work has been to use a simple, analytical procedure based on open literature data to determine the energy and emissions of almost all aspects of putting together an electric vehicle and operate it. These are weighed-in against GHG emissions due to conventional fossil fuel cars. This type of analysis becomes imperative as it is felt that global progress towards emissions control is stalling currently while it is expected to be accelerating.[7-9]

**Electric Vehicles (EVs) for transportation:**

Cars are a major part of road transportation, especially in the developed world. As per capita GDP increases, car ownership also increases showing a strong correlation between the two. The demand for passenger vehicles therefore has been predicted to increase considerably as a result of GDP growth in populous countries like China and India. In the Year 2018 alone ~ 95.6 Million vehicles were produced out-of-which ~ 73.7 % or 70.5 million vehicles were passenger cars.[10,11] The internal combustion engine (ICE) in these vehicles converts chemical energy of fossil fuels to mechanical energy which drives the vehicles. This mode of energy conversion and utilisation is inherently inefficient with only ~ 25 – 30 % efficiency and hence results in large amounts of energy as waste heat and GHG emissions. Electric motors on the other hand convert electrical energy into mechanical energy with an efficiency of at least ~ 75 % and hence offer an attractive alternative to ICE for automobiles which is the main reason for the International Energy Agency to push adoption of electric vehicles. In spite of this highly attractive feature, automobiles operated by electric motors have not made considerable in-roads. The total number of cars on the roads during the period 2012 to 2018 has risen steadily to ~ 481.4 Million while the cumulative number of battery using cars during the same period is only 3.27 Million,[12] 0.68 %, clearly showing that there are considerable factors which impede the growth of electric vehicles, EVs. The low penetration of EVs is also probably due to adverse media reports about various environmental effects of EVs and cars in general.[13,14] Before discussing the different materials intensity of EV, it is important to identify the energy/power requirement for typical driving conditions. Since the actual driving conditions depend on people and external factors such as roads and traffic, power requirement for near ideal conditions are considered here. The power requirement for two separate configurations, a suburban car with a smaller weight and cross-sectional area 1500 Kgs and 3 m$^2$ respectively and a typical sports utility vehicle (SUV) with a longer foot print and higher weight 6 m$^2$ and 2500 Kgs have been determined based on simple mechanics (**see Supplementary for details**). The power required to drive the vehicle decreases with decreasing weight of the vehicle and in turn reduces the emissions associated with manufacturing and running the vehicle.[15] Consider a typical non-stop driving at a constant speed of 75 Kmh$^{-1}$ and distance of 300 Km driving range before charging the EV. The energy required to overcome rolling resistance for this driving condition is ~ 12.3 KWh while that required to overcome air drag is ~ 23.5 KWh for the car with a frontal cross-sectional area of 3 m$^2$, a suburban car. The total energy required therefore will be 35.8 KWh under ideal conditions. The energy conversion processes on the other hand are not ideal and hence considering a loss of only 25 % during energy conversion from electrical energy to mechanical energy gives the energy requirement to be 44.8 KWh for the motor and subsequently energy stored in the battery. This translates to 149 Wh Km$^{-1}$ or 29.8 Wh Kg$^{-1}$ of car weight. If on the other hand the same journey is undertaken with a SUV, the energy required would be 97 KWh which will be equivalent to 324 Wh Km$^{-1}$ or 38.8 Wh Kg$^{-1}$. These results clearly show that the energy requirement scales with the distance travelled and speed at which the distance is covered for any type of car.

**Status and demand projections for EVs:**

The demand and growth of both Hybrid Electric Vehicles (HEV), Plugged-in Hybrid Electric Vehicles (PHEV) and All Electric Vehicles (AEV) depends on various factors with cost and government policies being the major factors. The actual number of EVs on the roads in the World till the Year 2018 is 3.27 Million[12] which has been predicted to increase significantly due to implementation of Intergovernmental Panel for Climate Change, IPCC guidelines for GHG emissions reduction strategy. Various governmental and non-governmental agencies and organisations have predicted the demand for EVs based on legislation in different countries and cost as factors. The IEA-Energy Technologies Perspective (ETP) report 2010 however bases its projections based on global emissions reduction goals as specified by IPCC which advocates 50 % reduction by the Year 2050 to restrict global warming to < 2 °C.[16] To achieve this objective two specific scenarios are considered – i) a scenario wherein all forms of EVs will coexist including the fuel cell based vehicles BLUE Map, and ii) majority of the vehicles will be AEVs with fuel cell based vehicles not making any contribution to the product mix BLUE Shift. BLUE Shift and BLUE Map are terms coined by the International Energy Agency. These terms probably signify recovering the clean, blue planet Earth by adopting emissions free technologies. According to these scenarios annual sale of EVs should increase steadily to reach ~ 157 million and 130 million respectively for the two scenarios. The detailed split between the different types of EVs is shown in Figures 1 (a) & (b). These staggering numbers clearly illustrate the importance of electric motors and batteries which are the two essential and critical components in an EV. There are different variants of electric motors that are currently being used/investigated for EVs. The most popular and extensively used variant however is the Permanent Magnet Synchronous Motor (PMSM) and hence the material intensity in this motor will be discussed. Since Li-ion batteries with the highest energy densities are the most preferred electrical energy storage option, materials intensity of these batteries will only be discussed.

**Permanent Magnet Synchronous Motor (PMSM):**

The motor for the drive train should be capable of delivering a constant high power at all speeds. The PMSMs have this capability compared to other motor variants and this makes them the preferred choice for all types of EVs, both in current and future versions. This brings into focus the important role of magnets in these motors which produce the required magnetic flux. Although different types of permanent magnets are available in the market, rare-earth elements based alloys, Nd-based alloy $Nd_2Fe_{14}B$ and its variants with a specific energy density of ~ 400 KJ m$^{-3}$, are the preferred choice of electric motor manufacturers. A complete materials intensity estimation for the electric motor should consider all the different elements but in this work materials intensity of Nd and Dy which are important constituents of permanent magnets will only be discussed as these are considered as 'critical' elements across nations. The amount of magnets used and the exact composition of the alloy are not exactly known as these are considered proprietary by the manufacturers. However based on the literature available the magnet weight has been found to vary between 1 Kg to 3.5 Kgs per vehicle and the Nd content in the magnets can vary between 21 wt.% to 32 wt.% while the Dy-content varies from 1 wt.% to 10 wt.%.[17-19] It should be noted here that there are several components in the vehicle which also require permanent magnets but are not really considered in this analysis as they are presumed to be insignificant to contribute to materials intensity. The requirement of rare-earth elements, Nd and Dy as per current EVs production and also as per BLUE Map & BLUE Shift predictions of IEA are given in Table 1 and Figure 2. These results show that if one considers the amount of metals consumed by the PMs in the Year 2018, low intensity values, as the basis, Nd- and Dy-metals requirement for 150 Million cars increases to ~ 7043 %, a staggering increase. This brings us directly into assessing the quantity of mineral ore that is required to be extracted to meet the requirement of Nd and Dy metals for the Year 2050.

The rare-earth elements are abundant and in fact orders of magnitude higher compared to precious metals like Au and they are as abundant as some of the transition metals such as Co, Ni, Cr, Nb and so on.[20] This fact is further confirmed by the respective Herfindahl-Hirschman Index, HHI value for Nd and Dy which is 9500 based on current production. However based on the reserves and future potential the HHI decreases to ~ 3100,[21] clearly indicating that unexploited reserves can be exploited in future. For the present study, the current major source of rare-earth elements for the World, the Bayan Obo mines in China, is considered and the mineralogical data of this ore body is used to estimate the status of reserves of Nd and Dy.[22,23] Assuming the precursor to extract the metals is oxides of the two metals, $Nd_2O_3$ and $Dy_2O_3$, the amount of oxides required to produce 1 Kg of the two metals can be estimated and is found to be 1.166 Kgs and 1.148 Kgs respectively. Using this as the basis and minerology of the ore, the exact amounts of ore required to produce Nd and Dy in the Year 2050 has been determined and is given in Table 2. The current level of production of rare-earth oxide concentrate is ~ 130,000 tonnes in 2018 which yields ~ 23,000 tonnes of Nd-metal and ~ 90 tonnes of Dy-metal. This level of production is certainly sufficient to meet the current EVs demand. However this production needs to be significantly ramped up if the 150 Million EVs requirement of 76.5 MT of oxide concentrate, has to be met. The mining and extraction of metals operations result in various products that are of consequence to the environment apart from producing the required metals. Hence it is important to estimate these energy and environmental parameters corresponding to Nd- and Dy-metals production[24] as per estimates given in Table 1 and these are given in Table 3. These results show that the energy and water requirement for production of the two important elements are significant in quantity and also processing results in release of emissions, both gases and dust into the atmosphere. The embodied energy and emissions given in Table 3 are for mining and extracting the rare-earth metals and does not include energy and emissions costs of making the exact alloy and subsequent manufacturing of the magnets which are not available in open, peer reviewed literature.

**<u>Batteries for EVs:</u>**

The second major component of all types of EVs is the battery which is the energy storage system and its capacity varies depending on the type and capacity of the EV. The battery capacity in HEVs & FCVs is small compared to PHEVs and AEVs as the main energy source in these vehicles is not the battery. As for batteries, there are several generations with the conventional Pb-acid battery having a low gravimetric energy density of ~ 25 Wh Kg$^{-1}$ compared to Li-ion batteries which have ~ 175 Wh Kg$^{-1}$.[25] Hence Li-ion batteries are more preferred energy storage devices in current and near future EVs and the materials intensity in these is limited to the requirement of Li as this element is mainly responsible for energy storage and transfer. The other elements, Co, Ni and Mn although are present, they are not considered critical. The amount of active material required per unit energy to be stored/delivered in a battery in general depends on the electrodes and electrochemical reactions.[26,27] A theoretical estimation of the minimum amount can be obtained using the Faraday's Laws and is expressed as;

$$I = Materials\ Intensity\ in\ g\ kWh^{-1} = {10^3 M}/{E_0 f C} \qquad (1)$$

where M is the molar mass in g mol$^{-1}$, $E_0$ is the electromotive force in V, f the fraction of material available for reaction and C the charge per mole. The above equation translates to $I = \left(65/f\right)$ for the case of Li assuming an electromotive force $E_0$ of 4 V. These batteries use a variety of electrodes such as LiCoO$_2$, LiMn$_2$O$_4$, LiFePO$_4$, LiNi$_{0.4}$Co$_{0.2}$Mn$_{0.4}$O$_2$ and so on and hence the Li-intensity varies from ~ 130 gkWh$^{-1}$ to ~ 173 gkWh$^{-1}$. A practical battery however requires more than the theoretical minimum to account for effects such as internal resistance of the cell and over-specification by the manufacturer.

The resulting Li-intensity therefore can vary from ~ 190 gkWh$^{-1}$ to 375 gkWh$^{-1}$.[28-31] The amount of Li required to make batteries with ~ 5 kWh rating used in HEVs and 40 kWh rating for PHEVs and AEVs is given in Table 4. It is seen that the amount of Li-required can vary between 0.143 MT to 2.25 MT depending on the energy storage capacity of the 150 million units. Although these quantities appear large, it is known that the World has sufficient reserves which can be exploited without having to include Li in the list of critical materials. The HHI value based on current levels of production therefore is 2900 which increases to 4200 based on reserves,[21] indicating that prospecting the available reserves has uncertainty either in terms of technology or economics.

The process of making the battery involves several steps starting from mining the required minerals to assembling the final battery pack with all the required connections and electronic components. All these processes require energy and they also have associated emission of gasses and particulate matter. An estimation of these environmental parameters has been performed for assembling a Li-ion battery having $LiNi_{0.33}Mn_{0.33}Mn_{0.33}O_2$ as the cathode, graphite as anode and $LiFP_6$ as the electrolyte. This combination of electrodes and electrolyte has been chosen as it is the most preferred battery chemistry by manufacturers. A cradle-to-gate impact analysis of a 1 kWh NMC111 battery manufacturing shows that it requires a total of 1125 MJ energy which has a weight of 7 Kgs.[32] Based on these energy and other environmentally relevant parameters, the energy required to make 5 kWh or 40 kWh battery packs is given in Table 5 together with all the environmental parameters. The energy required to make 2 Million batteries is ~ 25 TWh which increases to 1877 TWh while the associated $CO_2$-e emissions increases from 5.83 MT to 437 MT for making 150 Million batteries.

**Discussion:**

If one considers that the EVs and ICE vehicles have nearly identical bodies and electronic instrumentation, the main difference boils down to the electric motor with a battery for energy storage in an EV vis-à-vis the ICE with fossil fuel storage unit. The fuel storage unit however is a mere tank and forms an integral part of the vehicular structure. Hence the embodied energy and related emissions associated with manufacturing this unit become negligible and need not be considered. The liquid fuel production on the other hand has two distinct stages – crude oil production and subsequent refinement. The emissions from the conventional ICE vehicle therefore are due to burning of the fossil fuel during driving together with manufacturing of the fuels. It has been reported that crude oil production results in emission of 10.3 g $CO_2$-e MJ$^{-1}$ while the refining process to produce the fuels results in 7.3 g $CO_2$-e MJ$^{-1}$.[33,34] The typical density and calorific value of gasoline have been determined to be 0.745 Kgl$^{-1}$ and 42.7 MJ Kg$^{-1}$ respectively.[35] The $CO_2$ emission due to driving has been decreasing steadily since Year 2007 from 160 g Km$^{-1}$ to a current level of 100 g Km$^{-1}$ as shown in Figure 3 due to improvements in combustion technology and availability of cleaner liquid fuels as a result of various legislations.[36] Hence assuming an average value of 100 g $CO_2$ emissions for every Km driven by the car, and an average fuel efficiency of 15 Kml$^{-1}$ of gasoline across the world, the carbon intensity of driving an ICE car for 20000 Km per Year has been determined. The carbon intensity of fuel for 150 Million cars in the Year 2050 is found to be 112.03 MT while that due to burning of this fuel in ICE is found to be 300 MT, together adding to 412.03 MT of $CO_2$ emissions per Year. These emissions have to be compared with those associated with Nd- and Dy- metals production, battery assembly of EVs and emissions associated with electrical energy required for driving the cars. The embodied energy and emissions associated with cradle-to-gate of EVs production are given in Table 6 in a consolidated manner. The data regarding manufacturing of rare-earth magnets, electric motors and Li-ores extraction & processing are not available and hence are not included here. For the case of low metals intensity the emission is 389.5 MT of $CO_2$-e while that for high metals intensity it is 455.6 MT of $CO_2$-e. The electrical energy required to run the EVs for the same distance per year can be estimated using the earlier determined at wheel energy of 149 Wh Km$^{-1}$ and is found to be 447 TWh or $1.61 \times 10^{12}$ MJ for the Year 2050. The carbon intensity of generating this required quantity of electrical energy has been determined using the global average value[37] of 476 g $CO_2$ KWh$^{-1}$ and it is found to be 212.8 MT. Hence the net $CO_2$ emissions due to manufacturing and running EVs can be determined and varies between 602.3 MT and 668.4 MT. The emissions associated with manufacturing the EVs remain in the atmosphere even after the life of batteries and the emissions due to running the EVs are recurring in nature. These running emissions due to ICE vehicles is typically much longer than the battery life which is ~ 10 years. This means the batteries will require replacement after 10 years and recycling old batteries if possible, both processes contributing to further emissions. The present analysis is based on the following assumptions;

1. ICE technology remains static at the current level and hence the emissions due to driving will not reduce below 100 g $CO_2$ Km$^{-1}$.
2. The crude oil production and refining technologies will not undergo development and maintain a status-quo till Year 2050.
3. The rare-earth metals extraction, refining and manufacturing will not undergo developments to reduce their carbon intensity.
4. The energy and emissions intensive battery making processes will not change and maintain current levels of intensity through the Year 2050.

An additional point that should be considered in evaluating the potential of EVs is the electrical energy generating plants as well as the mechanisms to re-charge the batteries. The net electrical energy requirement of the EVs is 447 TWh whereas a typical power plant of 600 MW capacity can generate only 5.256 TWh even if it runs without disruptions through the year. This implies that about 85 power plants will be required exclusively to supply the energy required for running the EVs through the year. The embodied energy and associated equivalent emissions involved in constructing these additional power plants will further add to the environmental cost of EVs.

The batteries in AEV have high capacity and unlike HEV and PHEV, the AEV does not have any source of energy on the go to recharge and hence they need to be recharged offline. Recharging these high powered batteries requires specialized equipment as it requires high energy transfer rates so that the time can be minimized. This aspect of recharging becomes apparent if one considers filling gasoline in conventional ICE driven vehicles. Typically, 50 l of gasoline with an energy content of ~ 0.5 MWh and a range of 750 Km can be filled into the tank in < 5 minutes. To achieve a similar energy transfer rate for recharging the Li-ion battery, the charging station should be working at a D.C. current of 150 A and 40 KV voltage, which is an extremely dangerous/unsafe practise. As a result several of the charging schemes take several hours for full recharge so that safety concerns can be maintained. A typical commercially available charging station has an A.C. input power rating of 100 to 200 KVA with an associated D.C. power output of 50 to 180 KW.[38] The charging equipment including the power cables therefore have to be specially designed both to deliver the required power and to meet the safety standards. The battery manufacturing processes which are the main contributors to emissions is being constantly researched and developed with alternate electrodes as well as battery chemistries.[39,40] These alternatives should result in reducing the energy and emissions intensity of the battery in the future. Parallely however, the ICE technology is also witnessing changes in the form of alternate gaseous and liquid fuels such as propane, compressed natural gas, hydrogen and methanol, ethanol, respectively.[41] The feed stock for most of these alternate fuels is conventional non-renewable sources such as natural gas, petroleum or coal while for ethanol it is essentially corn or sugar cane or cellulosic biomass. The main problem with ethanol or biodiesel is that it requires agricultural land and water as feed stock which interferes into the food chain. All these fuel substitutes are anticipated to further reduce the emissions from ICE vehicles as they are relatively 'clean' and thus making it even harder for EVs to compete. These results clearly show that the emissions from EVs are at least 1.5 times higher compared to those from conventional ICE driven cars and that the objective of GHG emissions reduction by substituting ICE cars with EVs cannot be realized using the current practises. Recently, a Shared Socioeconomic Pathways model was used to analyse the role of EVs in the US alone and it was found that substitution of ICE vehicles with EVs will not bridge the $CO_2$ emissions gap.[42] Both technological as well as governing policies have to change significantly in order to meet the goal of reducing emissions so that GMST variations are kept < 2 °C as per various recommendations.

**Conclusions:**

In Summary, electrification of transportation, specifically cars is being promoted by several countries as well as organisations across the world as a means to decarbonise the sector and thus reduce global emissions to limit warming to < 2 ˚C. In principle, this is a good concept to follow so as to bring the global warming phenomenon under control. A near complete cradle to gate analysis of this concept based on the current status of different technologies and hence materials needed for operationalizing this concept however clearly shows that the primary objective of reducing emissions cannot be achieved. The $CO_2$ emissions due to 150 million electric cars running in the Year 2050 will be almost 2 times higher compared to emissions due to fossil fuel based cars which is ~ 412 MT. The embodied energy and emissions of EVs scale linearly with the number of cars, clearly showing that unless there is a first order change in battery making, it can never compete with ICE vehicles. This is because the Li-ion battery making process is extremely energy and carbon intense requiring nearly 1500 TWh of energy and emitting 379 MT of $CO_2$-e gases constituting about 50 % of total emissions due to EVs. The current electric motor and battery designs depend critically on high supply chain risk elements such as Nd, Dy, Co and Ni and developments on these fronts to substitute these with low risk, environmentally friendly elements will contribute significantly to truly justify promoting the EVs. Since the ICE vehicles are also witnessing considerable technological progress both in terms of engine design as well as developing cleaner fuels including gases, a possible best case scenario would be to have a combination of both technologies, i.e. PHEV instead of AEV for transportation.

Another alternative for the use of EVs is in mass transport systems such as buses and long haul freight transport. Utilization of these EVs as mass public transport systems with the last mile connectivity provided by electric bicycles/bikes/tricycles can become very effective in combating the emissions. This will reduce drastically the total number EVs required and will also result in extremely high utilization of these vehicles compared to individually owned vehicles. This strategy however will require socio-political acceptance and policy frameworks. This model can be an effective solution for low income and developing countries wherein the cost of owning an EV is still too large compared to GDP. The electricity requirement and infrastructural requirements required for operating the EVs however will still need to be factored. Making the primary electricity generation processes to become completely green will also reduce emissions intensity of EVs. Towards this goal, significant efforts are being made across the world to produce electricity from renewable sources such as Sun, wind, oceans, biomass and so on. These technologies are anticipated to reduce the emissions loading of atmosphere compared to electricity production using conventional fuels. It should however be noted that these technologies will only reduce but not reverse the emissions, i.e. sequester the existing emissions, which is required to halt the global warming process.

**Acknowledgements:** The author wishes to acknowledge IIT Bombay for the provision of infrastructure facilities to carry out this work.

The author declares that this work did not receive any funding and that there are no conflicts of interest.

**Tables:**

|  | Low Intensity | High Intensity |
|---|---|---|
| **PM weight** | 1 Kg | 3.6 Kgs |
| **Nd-content, wt.%** | 21 | 32 |
| **Dy-content, wt.%** | 1 | 10 |
| **For the Year 2018; 2.1 Million EVs** | | |
| **Nd-metal, Tonnes** | 529.2 | 2,903.1 |
| **Dy-metal, Tonnes** | 25.2 | 907.2 |
| **For the Year 2050; 150 Million EVs** | | |
| **Nd-metal, Tonnes** | 37,800 | 207,360 |
| **Dy-metal, Tonnes** | 1,800 | 64,800 |

**Table 1:** The intensity of the two rare-earth metals Nd and Dy, i.e. the actual amounts required to make the permanent magnets in EV motors, in the two limits low and high, for the two years 2018 and 2050. It shows that requirement of both the metals increases by ~ 71 X in 2050 compared to that in 2018, a phenomenal increase necessitating prospecting huge amount of ore. The error associated with the estimated metal values is about ± 10 % due to possible variations in ore concentration and extraction efficiency.

| Metal | Ore requirement, Million Tonnes | |
|---|---|---|
| | Low Intensity | High Intensity |
| Nd-metal | 3.981 | 21.838 |
| Dy-metal | 47.447 | 1708.1 |

**Table 2:** The amount of ore required to produce Nd and Dy metals in the Year 2050, 150 million EVs, determined using mineralogical data of current majority producer, Bayan Obo mines in China, is given here. The amount of ore required to be processed to extract the required amount of Dy is far greater than that for Nd and since the ore contains both metals, mining the required amount of ore for Dy should result in producing the Nd metal.

|  | Neodymium, Nd | | Dysprosium, Dy | |
|---|---|---|---|---|
|  | Low Intensity | High Intensity | Low Intensity | High Intensity |
| **Energy required, TWh** | 14.74 | 80.87 | 1.86 | 66.73 |
| **Water required, Gl** | 4.46 | 24.47 | 0.25 | 8.81 |
| **Emissions:** | | | | |
| $CO_2$, MT | 10.93 | 59.93 | 1.28 | 45.94 |
| CO, T | 29,673 | 162,778 | 1,224 | 44,064 |
| $SO_2$, T | 88,074 | 483,149 | 11,682 | 420,552 |
| $NO_x$, T | 33,264 | 182,447 | 4,158 | 149,688 |
| $CH_4$, T | 43,848 | 240,538 | 4,734 | 170,424 |
| **Dust Particles, T** | 167,076 | 916,531 | 22,248 | 800,928 |

**Table 3:** The mining, processing and extraction of rare-earth metals Nd and Dy requires not only energy and water but also results in environmental impact assessed by emission of various GHGs. These are given here based on typical data corresponding to production using ore of Bayan Obo mines in China. Apart from the gas emissions, these operations also result in emitting huge quantities of dust into the atmosphere. These values correspond to producing 150 million electric motors for EVs. The error in the emissions is about $\pm$ 10 % depending on the exact processes used for extraction and metal recovery.

| Battery Capacity, KWh | 5 | | 40 | |
|---|---|---|---|---|
| | Low | High | Low | High |
| Li-metal, MT | 0.1425 | 0.2786 | 1.14 | 2.25 |

**Table 4:** The amount of Li-metal required to make batteries of 5 kWh and 40 kWh energy which can be used in HEVs, PHEVs and AEVs in the two limits, low and high, for the Year 2050, 150 million EVs are given. The low limit corresponds to 190 gkWh$^{-1}$ while the high limit is 375 gkWh$^{-1}$ of Li. It should be noted that these battery energies are typical of suburban vehicles and not SUVs which need even higher energy batteries.

|  | 5 KWh | 40 KWh |
|---|---|---|
| **Total Energy Required / Embodied Energy, TWh** | 234.6 | 1,876.7 |
| **Water required, Gl** | 564 | 4,512 |
| **Emissions:** | | |
| **$CO_2$-e, MT** | 54.7 | 437.4 |
| **$SO_x$, MT** | 0.6 | 4.8 |
| **$NO_x$, T** | 72,675 | 581,400 |
| **PM10 Dust, T** | 35,925 | 287,400 |

**Table 5:** The total amount of energy and water required to make the battery pack that drives the EVs and the consequent emissions during the battery manufacturing process are given here. These data are for a 40 kWh energy battery pack which will be used in suburban PHEVs and AEVs and can have a longer driving range. The values given correspond to assembling 150 million units. The error in the energy and emissions values is about ± 10 %.

|  | Nd-production | Dy-production | Battery pack assembly |
|---|---|---|---|
| **Embodied energy, TWh** | 14.74/80.87 | 1.86/66.73 | 1,876.7 |
| **Water, Gl** | 4.46/24.47 | 0.25/8.81 | 4,512 |
| **$CO_2$, MT** | 10.93/59.93 | 1.28/45.94 | - |
| **$CO_2$-e, MT** | - | - | 437.4 |
| **CO, T** | 29,673/162,778 | 1,224/44,064 | - |
| **$SO_x$, T** | 88,074/483,149 | 11,682/420,552 | 4.8 MT |
| **$NO_x$, T** | 33,264/182,477 | 4,158/149,688 | 581,400 |
| **Dust Particles, T** | 167,076/916,531 | 22,248/800,928 | 287,400 |

**Table 6:** The consolidated energy, water and GHG emissions for Nd-, Dy- metals production from the ores and assembly of 40 KWh NMC111 Li-ion battery pack. The data for rare-earth metals production in both the low and high intensity requirements are given here. It is seen that the battery assembly is the dominant contributor to both energy consumption and environmental impact. The error in these values is about $\pm$ 10 % as mentioned in previous tables.

**Figures:**

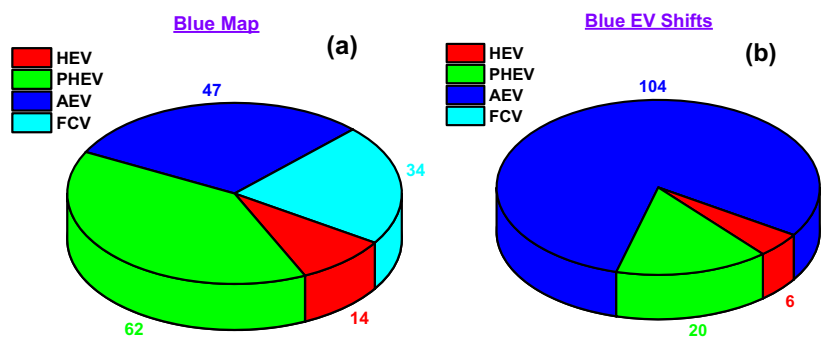

**Figure 1:** The fractional make-up of different type of electric vehicles, EVs that will be required to limit global warming to < 2 °C by the Year 2050 as per IPCC recommendations is shown. The fuel cell electric vehicles FCV according to Blue EV Shift do not contribute to the mix while they contribute ~ 21.6 % as per Blue map. It is seen that having predominantly AEVs reduces the total number of vehicles required to realise the same reduction in emissions. The numbers given are in Millions.

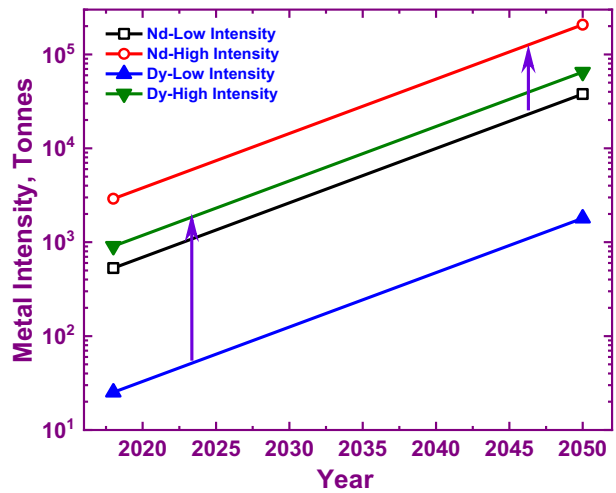

**Figure 2:** The amount of Nd and Dy metals required to increase the number of electric vehicles, EVs from 2.1 Million in the Year 2018 to 150 Million in the Year 2050 increases significantly. Note the metal intensity is shown in logarithmic scale. These estimates are based on the two assumptions; i) processing the metal ingots into finished alloy magnet of the required size and shape results in a material waste of 20 %, and ii) the market penetration of PMSM for EVs is 90 % in the absence of alternative non-magnet based motor designs.

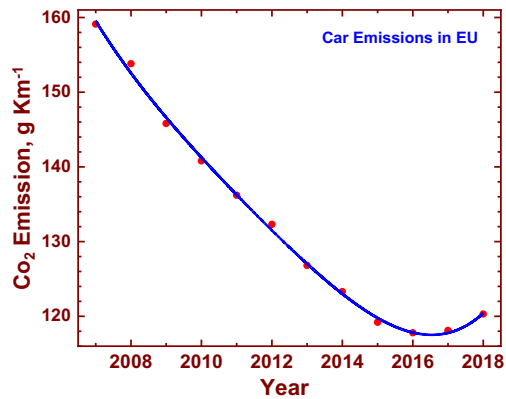

**Figure 3:** The $CO_2$ emissions from conventional internal combustion engine vehicles, cars, in the European Union have been steadily declining due to refinements in IC engines and availability of clean fuels due to regulations implemented across countries. The small rise seen after the Year 2017 is within the emission limits and not real. With the implementation of EU-6 and equivalent regulations across the world this is expected to decrease further to < 100 g Km$^{-1}$. These general trend of falling emissions has been observed across the world.

Supplementary Information for

**Electric cars – assessment of 'green' nature vis-à-vis conventional fuel driven cars**

Satish Vitta[*]

Department of Metallurgical Engineering and Materials Science

Indian Institute of Technology Bombay

Mumbai 400076; India.

The electric vehicles that are currently being produced are of different variety and all of them are not truly electric. Hence the different types are clearly described together with the nature of drive train present in these vehicles so that the exact electric motor and battery requirement can be identified.

Electric Vehicles (EVs) can be broadly classified into 4 different types: 1) Hybrid Electric Vehicle (HEV), 2) Plug-in Hybrid Electric Vehicle (PHEV), 3) All Electric Vehicle (AEV), and 4) Fuel Cell Vehicle (FCV).[1] Among these, the FCV which has hydrogen gas as energy source will have a small battery to act as an intermediary and since the maturity level of this technology is not at an advanced stage compared to other 3 variants, it is not discussed here in detail.

1) Hybrid Electric Vehicles (HEV): As the name indicates these vehicles are driven both by conventional ICE as well as a battery powered electric motor with the two drive trains configured to operate either in parallel or in series. The main energy source however is conventional fossil fuel, oil, with the battery getting charged by the ICE and acting as a mere store house of energy. The battery therefore cannot be charged using an external power source and gets charged by a small generator/motor which is connected to the ICE. Since the battery in this vehicle merely stores a small amount of energy its capacity is very small. The vehicle therefore can travel only very short distances at low speeds using electricity from the battery. In these vehicles the electric option is rather notional and it contributes solely to enhance the conventional fuel economy or the range of the vehicle.

2) Plug-in Hybrid Electric Vehicles (PHEV): These vehicles on the other hand have a single drive train driven by the electric motor. This configuration can also be called a series hybrid configuration with the conventional ICE being coupled to the battery via a generator. Since the ICE mainly generates power through a small generator while the driving is done by electric motor, this configuration requires a larger battery for power storage compared to HEV. The main advantage however is that the battery can also be charged by plugging to conventional electrical power. The electrical range for a PHEV is much larger than the HEV but far less compared to a full blown AEV and can range between 10 Km to 50 Km. Depending on the configuration and architecture of the drive train the PHEV can work either in the electric mode or in the combination mode depending on the vehicle speed/power requirement.

3) All Electric Vehicle (AEV): An AEV has a single drive train driven by the electric motor which receives energy from the battery. Since the system is driven solely by electric power from the battery, the battery capacity is much larger compared to HEV. As it does not have a ICE to charge the battery 'on the go', it can only be charged when it is stationary by sourcing electrical power from external sources. The range of these vehicles currently is therefore limited by the capacity of the battery to store electrical energy.

**Mechanics of motion:**

The efficiency of engines/motors i.e. energy conversion into mechanical energy is relatively constant and does not change significantly with driving conditions with the fossil fuels based internal combustion engine ICE having an efficiency of ~ 25 % while the electric motor can have an efficiency of ~ 80 %. The vehicle performance at best therefore will follow the guidelines given below which will be the maximum limit.

The energy consumption in a vehicle/car has four distinct components;[2,3]

1. Accelerating to designated speed and decelerating to a halt;
2. Rolling resistance due to a combination of factors such as friction between the tyres & the road, energy absorbed by various components and the energy transferred to the ground;
3. Energy required to overcome the 'air drag'; and
4. Energy efficiency of the engine which converts chemical/electrical energy into mechanical energy.

If the distance d between starting point and stopping point is large i.e. highway driving and not city driving, the power lost in the braking process given by; $1/2 \left( m_c v^3 / d \right)$ where $m_c$ is the mass of the car and v the velocity, can be neglected as it is inversely proportional to the distance travelled d. This leaves the two main contenders of energy consumption that contribute to energetics of a car – the rolling resistance and air drag which depends on v.

The rolling resistance is given by the equation;

$$Rolling\ resistance\ Power = P_r = (C_r \times m_c \times g \times v) \qquad (1)$$

where $C_r$ is the coefficient of rolling resistance and g the acceleration due to gravity. The power required to overcome air drag is given by the relation;

$$Air\ drag\ Power = P_d = \frac{1}{2}(\rho_{air} \times C_d \times A_{car} \times v^3) \qquad (2)$$

where $C_d$ is the drag coefficient, $\rho_{air}$ the air density and $A_{car}$ the frontal cross-sectional area of the car. The above equations show that the power required to overcome rolling resistance is directly proportional to the vehicle speed v while the air drag has a cubic dependence on speed v, which are shown in Figure S1 for two different car configurations.

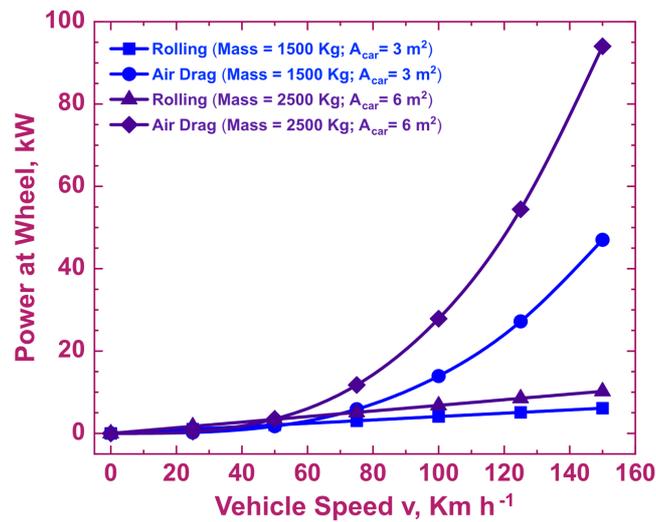

**Figure S1:** The at wheel power requirement determined using equations (1) & (2) for two cases, a sub-urban compact vehicle with lower weight and cross-sectional area and a sports utility vehicle (SUV) with higher weight and cross-sectional area is shown as a function of vehicle speed v. The power required to overcome rolling resistance as well as air drag increases with increasing v with air drag becoming dominant at high v.

As the vehicle weight or cross-sectional area increases, the speed at which it becomes dominant decreases. The above data is used to determine the power requirement of electric vehicles with either a sub-urban or a sports utility configuration.

**Metals Intensity and Environmental parameters determination:**

The methodology used to determine materials intensity, embodied energy and emissions given in Tables 1-6 is given below.

The amount of permanent magnets used in the electric motor under the two scenarios – low and high intensity are given in Table 1. The exact composition of the different elements in the permanent magnets taken from open literature shows that it can vary between 21 wt.% and 32 wt.% Nd and 1 wt.% and 10 wt.% Dy.[4-6] Hence the low intensity case 1 Kg magnet in the motor with the alloy having 21 wt.% Nd & 1 wt.% Dy and high intensity case – 3.6 Kgs of magnet with 32 wt.% Nd & 10 wt.% Dy are used as the two limiting cases.

Currently, the major supplier of rare-earth elements to the world is China and the minerals required to extract these elements are concentrated in Bayan Obo mines. Hence the mineralogical and processing data of these minerals is used to determine the ore requirement, amount of ore required to be mined.[7,8] The cumulative rare-earth oxide content in these ores is found to be 6.22 wt.% which is generally termed as REO concentrate. The amount of $Nd_2O_3$ in this REO concentrate is 17.8 wt.% while that of $Dy_2O_3$ is only 0.07 wt.%. This information is used to determine the exact amount of ore required to be mined to extract the metals that go into the permanent magnet alloys.

Again, since the Bayan Obo mines are the chief sources of rare-earth elements, the processing parameters corresponding to these ores have been used to determine the environmental impact and they are given below;[9]

|  | Nd, per Kg | Dy, Per Kg |
|---|---|---|
| **Primary Energy, MJ** | 1404 | 3707 |
| **Process Water, Kg** | 118 | 136 |
| **$CO_2$, Kg** | 289 | 709 |
| **CO, Kg** | 0.785 | 0.68 |
| **$SO_2$, Kg** | 2.33 | 6.49 |
| **$NO_x$, Kg** | 0.88 | 2.31 |
| **$CH_4$, Kg** | 1.16 | 2.63 |
| **Dust particles in air, Kg** | 4.42 | 12.36 |

The absolute amount of Li present in the battery depends on its chemistry which in turn depends on the electrodes used. This information is considered to be proprietary by battery manufacturers and hence is not easily available in open literature. Also, different manufacturers have different methods of accounting for over voltage, safety and so on. As a result the typical Li content in the battery is found to vary between the limits of 190 g $KWh^{-1}$ and 375 g $KWh^{-1}$. Hence these values are used to determine Li-metal intensity.[10-13]

The cradle-to-gate impact of assembling a 1 KWh NMC111, $LiNi_{0.33}Mn_{0.33}Co_{0.33}O_3$ battery pack, assessed from large scale industrial production of Li-ion batteries has been used to determine the environmental impact. The different environmental parameters are;[14]

|  | Per KWh NMC111 battery |
|---|---|
| **Embodied energy, MJ** | 1126 |
| **Process water, l** | 752 |
| **$CO_2$-e, Kg** | 72.9 |
| **$SO_x$, g** | 800 |
| **$NO_x$, g** | 96.9 |
| **PM 10 in air, g** | 47.9 |